\documentclass[a4paper,reqno,11pt]{amsart}
\usepackage{amsmath}
\usepackage{amssymb}
\usepackage{cases}
\usepackage{enumitem}
\allowdisplaybreaks[4]

\newtheorem{theorem}{Thm}[section]
\newtheorem{lemma}[theorem]{Lemma}
\newtheorem{definition}[theorem]{Def}

\usepackage{todonotes}

\begin{document}

\title{General Rational Solutions and Soliton Solutions of the Nonlocal Resonant Nonlinear Schrödinger Equations}
\date{}

\author{Bo Wei, Zhenyun Qin, Gui Mu}
\address{Zhenyun Qin, School of Mathematical Sciences, Fudan University, Shanghai, 200433, China}
\email{zyqin@fudan.edu.cn}

\begin{abstract}
General rational solutions for the nonlocal resonant nonlinear Schrödinger equations are derived by
using the Hirota's bilinear method and the KP hierarchy reduction method. These rational solutions are presented in
terms of determinants in which the elements are algebraic expressions.
A weaker condition is given for KP reduction in the nonlocal case.
The dynamics of first-order solutions are investigated in details.
As a special case, we studied rantional solutions of the RNLS equation. 
Moreover, soliton solutions of the RNLS equation are given by using the Bäcklund transformation and nonlinear superposition formula.
\end{abstract}


\maketitle
\pagestyle{myheadings} \markboth{\hfill Bo Wei, Zhenyun Qin, Gui Mu\hfill}{}

\tableofcontents

\section{Introduction}
For most nonlinear wave phenomena, two types of model equations can be derived. One is the so-called "long-wave model," represented 
by the Korteweg-de Vries (KdV) equation. The other is the "short-wave model," represented by the nonlinear Schrödinger equation 
(NLS equation). The NLS equation serves as a model for many nonlinear systems and finds applications in various fields such as fluid 
mechanics, nonlinear optics, nonlinear acoustics, and solid-state thermal pulses. It can be divided into two cases: the focusing NLS 
equation and the defocusing NLS equation. The resonant NLS equation (RNLS equation)
\begin{equation}\label{RNLS}
    \begin{aligned}
        i\Psi_t+\Psi_{xx}-\frac{1}{2}|\Psi|^2\Psi=s\frac{|\Psi|_{xx}}{|\Psi|}\Psi
    \end{aligned}
\end{equation}
is a type of NLS equation that can be considered 
as a third kind of NLS equation between the focusing and defocusing cases. It is integrable and exhibits resonant solitary wave phenomena 
in certain situations \cite{ref5}. The origin of the RNLS equation can be traced back to the quantum potential proposed by the French 
mathematician de Broglie in 1926, and the RNLS equation can be regarded as a special case of the NLS equation describing the propagation 
of solitons in this quantum potential. Subsequently, Bohm (1952), Salesi (1996), and others applied quantum potential in the fields of 
quantum mechanics and stochastic mechanics. In 2002, Lee and Pashaev formally established the RNLS equation \cite{ref1} and applied it in 2006 
to construct a propagation model for one-dimensional nonlinear magnetoacoustic waves in collisionless plasma under the influence of a 
transverse magnetic field \cite{ref2}.

There has been extensive research on the RNLS equation, including stability and modulation instability \cite{ref7}-\cite{ref8}. 
Moreover, significant progress has been made in obtaining soliton solutions for various types of RNLS equations. For example, 
nontrivial boundary condition solitons for the RNLS equation \cite{ref5}, generalized solutions for the RNLS equation \cite{ref15}, 
optimal solitons for integrated RNLS equations and RNLS equations with arbitrary indices \cite{ref16}-\cite{ref17}, non-self-controllable 
solitons for variable coefficient RNLS equations \cite{ref18}-\cite{ref14}, RNLS equations with Kerr nonlinearity \cite{ref9}, RNLS equations 
with power nonlinearity \cite{ref10}, RNLS equations with parabolic nonlinearity \cite{ref11}-\cite{ref12}, 
and fully nonlinear RNLS equations \cite{ref13}. The process of solving these equations using the Hirota bilinear method \cite{ref5} is 
the most direct. In addition to soliton solutions, the RNLS equation also admits multi-rogue wave solutions and lump solutions \cite{ref3}.

Some soliton equations possess the property that the evolution of their solutions depends not only on time and spatial coordinates 
but also on nonlocal interactions. Such soliton equations are referred to as nonlocal. Ablowitz and Musslimani were among the first 
to study nonlocal equations, focusing on nonlocal Schrödinger equations \cite{ref39}. Since then, nonlocal equations have received increasing 
attention and have produced abundant results, such as nonlocal DS $\textnormal{\uppercase\expandafter{\romannumeral1}}$ equations and 
(2+1)-dimensional nonlocal Schrödinger equations \cite{ref40}, nonlocal mKdV equations, and nonlocal SG equations \cite{ref41}. 
Related results regarding soliton solutions \cite{ref4}, rogue wave solutions, 
and lump solutions \cite{ref3} have been obtained.
This paper focuses on the nonlocal RNLS equation
\begin{equation}\label{NRNLS}
    \begin{aligned}
        &i\Psi_t+\Psi_{xx}-\frac{1}{2}h\Psi=s\frac{|\Psi|_{xx}}{|\Psi|}\Psi,\\
        &h=\int_0^1|\Psi(x,t,\xi)|^2d\xi,
    \end{aligned}
\end{equation}

This paper primarily adopts the KP reduction method. The Kadomtsev-Petviashvili equation (KP equation) is a high-dimensional extension 
of the Korteweg-de Vries equation (KdV equation). Sato discovered that the polynomial solutions of the bilinear KP equation are essentially 
equivalent to the characteristic polynomials of the general linear group \cite{ref35}, which highlights the importance of the KP equation 
in soliton equations. Later, Sato established the Lax pairs for the KP hierarchy using pseudo-differential operators, revealing the 
equivalence between the KP equation and the motion of a point on the Grassmann manifold \cite{ref35}, where the Plücker relations on the 
Grassmann manifold precisely correspond to the bilinear KP equation. In 1983, Jimbo and Miwa studied the algebraic structure of soliton 
equations from the perspective of group theory and established the connection between infinite-dimensional Lie algebras and representations 
on function spaces of soliton equations \cite{ref20}. The corresponding Lie algebra for the KP equation is $gl(\infty)$, thus occupying a central 
position in soliton equations. Hirota conducted related work on the structure of the bilinear KP equation and soliton equations \cite{ref35}.

Satsuma, Freeman, Nimmo, and others discovered that the soliton solutions of the KP equation can be expressed using Wronskian determinants, 
where the bilinear KP equation becomes an identity involving determinants. Subsequently, Hirota used Gram determinants to express the 
soliton solutions of equations in the KP hierarchy and showed that the bilinear equations in the KP hierarchy correspond to Pfaffian 
identities \cite{ref35}. The work of Sato, Jimbo, Miwa, and others provided rational solutions for equations in the KP hierarchy. 
Following that, the KP reduction method has been employed in similar studies on various equations, such as rational solutions for 
the nonlinear Schrödinger equation \cite{ref22}, the Yajima-Oikawa equation \cite{ref23}, the Boussinesq equation \cite{ref24}, 
the Schrödinger-Boussinesq equation \cite{ref25}, the Davey-Stewartson $\textnormal{\uppercase\expandafter{\romannumeral1}}$ equation 
\cite{ref26}, rational solutions for (2+1)-dimensional dispersion long wave equations \cite{ref27}, and more. These results encompass 
different types of rational solutions, including rogue wave solutions, lump solutions, and breather solutions.

For any parameter $s$, the RNLS equation $(\ref{NRNLS})$ is an integrable system. However, the equation exhibits different properties 
depending on the value of $s$ \cite{ref5}. When $s<1$, the RNLS equation can be reduced to the traditional nonlinear Schrödinger 
equation, while when $s>1$, the equation exhibits resonance solitonic phenomena and is called the resonant nonlinear Schrödinger 
equation. This paper only considers the case where $s>1$ and sets $s=1+\beta^2(\beta\neq 0),$
then the equation can be reduced to a reaction-diffusion system (RD system)
\begin{equation}
    \begin{aligned}\label{NRD}
        -&e^+_t+e^+_{xx}=-\frac{1}{2\beta^2}he^+,\\
        &e^-_t+e^-_{xx}=-\frac{1}{2\beta^2}he^-,\\
        &h=-\int_0^1e^+e^-d\xi
    \end{aligned}
\end{equation}
by Madelung transformation
\begin{equation}\label{Madelung}
    \begin{aligned}
        \Psi=\sqrt{\rho}e^{-iS},
    \end{aligned}
\end{equation}
where
\begin{equation}\label{epm}
    \begin{aligned}
        e^+=\sqrt{\rho}e^{\frac{S}{\beta}},\quad e^-=-\sqrt{\rho}e^{-\frac{S}{\beta}}.
    \end{aligned}
\end{equation}

Take $\beta=\frac{1}{2}$. Due to the complexity of integrating, in practical solving of equations, we replace the integral term with 
discrete summation, i.e., consider
\begin{equation}\label{LNRD}
    \begin{aligned}
        -&e^+_{k,t}+e^+_{k,xx}=-2he^+_k,\\
        &e^-_{k,t}+e^-_{k,xx}=-2he^-_k,\\
        &h=-\sum\limits_{k=1}^Le^+_ke^-_k.
    \end{aligned}
\end{equation}

The Bäcklund transformation is also one of the important methods for solving soliton equations. In 1883, Bäcklund first used this 
method to solve the SG equation\cite{ref35}, and it was named after him. Its characteristic is to transform the original equation 
into a new equation using a known solution, which allows obtaining many solutions of the equation through purely algebraic operations. 
At the same time, the Bäcklund transformation satisfies the nonlinear superposition formula, which can significantly reduce the 
difficulty of solving the new equation\cite{ref34}. 
Hirota proposed the Bäcklund transformation and nonlinear superposition formula for bilinear equations using bilinear operators and studied 
the bilinear Bäcklund transformation of KdV-type nonlinear equations\cite{ref35}. Subsequent studies on the Bäcklund transformation of 
nonlinear equations have adopted the Hirota bilinear method, such as the KdV equation\cite{ref35}, Boussinesq eqaution\cite{ref28}, 
NLS eqaution\cite{ref29}, Benjamin-Ono eqaution\cite{ref34}, 
BKP eqaution\cite{ref36}, Genralized KP eqaution\cite{ref21}-\cite{ref38}, shallow water wave eqaution\cite{ref30}, 
the coupled Ramani eqaution\cite{ref31}, Nizhnik-Novikov-Veselov eqaution\cite{ref32}, etc. 
In this paper we use Bäcklund transformation to obtain soliton solutions of the RNLS eqaution.

\section{Gram determinant solutions to a bilinear equation}
In this section we investigate a bilinear equation
\begin{equation}\label{nengsuanshuangxianxing}
    \begin{aligned}
        &(D_x^2-2r_kD_x-D_t)G_k^+\cdot F=0,\\
        &(D_x^2+2r_kD_x+D_t)G_k^-\cdot F=0,\\
        &-D_xD_yF\cdot F=2\sum\limits_{k=1}^L (G_k^+G_k^--F^2),
    \end{aligned}
\end{equation}
where the Hirota operator is a bilinear operator defined on functions, given by
\begin{equation}
    \begin{aligned}
    \notag
    &p(D_x,D_y,\cdots ,D_t)a(x,y,\cdots ,t)\cdot b(x,y,\cdots ,t)\\
    =&p(\partial_x-\partial_{x^{\prime}},\partial_y-\partial_{y^{\prime}},\cdots \partial_t-\partial_{t^{\prime}})
    a(x,y,\cdots ,t)b(x^{\prime},y^{\prime},\cdots ,t^{\prime})|_{x^{\prime}=x,y^{\prime}=y,\cdots ,t^{\prime}=t}.
    \end{aligned}
\end{equation}

To state the theorem, we define several symbols: $(n)$ denotes an L-element array $(n^{(1)},\ n^{(2)},\cdots ,\ n^{(L)})$, 
with length $l(n)=n^{(1)}+n^{(2)}+\cdots +n^{(L)}$. For the array $(n)$, we define $(n^{(k)}+s)$ as 
$(n^{(1)},\ n^{(2)},\cdots ,\ n^{(k)}+s,\cdots ,\ n^{(L)})$, where 0 denotes a zero array.

\begin{theorem}\label{LNgram}
    Let $m_{ij}^{(n)},\ \phi_i^{(n)},\ \psi_j^{(n)}$be functions of $x_1,\ x_{-1},\ x_2$ $(1\leq i,j\leq N)$, where $(n)=(n^{(1)},\cdots ,n^{(L)})$, 
    and $n^{(k)}\in \mathbb{Z}$, such that
    \begin{equation}
        \begin{aligned}
            \notag
            &\partial_{x_1}m_{ij}^{(n)}=\phi_i^{(n)}\psi_j^{(n)},\\
            &\partial_{x_2}m_{ij}^{(n)}=(\partial_{x_1}\phi_i^{(n)})\psi_j^{(n)}-\phi_i^{(n)}(\partial_{x_1}\psi_j^{(n)}),\\
            &(\partial_{x_2}+2r_k\partial_{x_1})m_{ij}^{(n)}=\phi_i^{(n^{(k)}+1)}\psi_j^{(n)}
            +\phi_i^{(n)}\psi_j^{(n^{(k)}-1)},\quad  k=1,\cdots ,L,\\
            &\partial_{x_{-1}}m_{ij}^{(n)}=-\sum\limits_{k=1}^L\phi_i^{(n^{(k)}-1)}\psi_j^{(n^{(k)}+1)},\\
        \end{aligned}
    \end{equation}
    \begin{equation}
        \begin{aligned}
            \notag
            &m_{ij}^{(n^{(k)}+1)}=m_{ij}^{(n)}+\phi_i^{(n)}\psi_j^{(n^{(k)}+1)},\quad  k=1,\cdots ,L,\\
            &\partial_{x_1}\phi_i^{(n)}=\phi_i^{(n^{(k)}+1)}-r_k\phi_i^{(n)},\quad  k=1,\cdots ,L,\\
            &\partial_{x_1}\psi_j^{(n)}=-\psi_j^{(n^{(k)}-1)}+r_k\psi_j^{(n)},\quad  k=1,\cdots ,L,\\
            &\partial_{x_2}\phi_i^{(n)}=\partial_{x_1}^2\phi_i^{(n)},\quad  \partial_{x_2}\psi_j^{(n)}=-\partial_{x_1}^2\psi_j^{(n)},\\
            &\partial_{x_{-1}}\phi_i^{(n)}=\sum\limits_{k=1}^L\phi_i^{(n^{(k)}-1)},\quad
            \partial_{x_{-1}}\psi_j^{(n)}=-\sum\limits_{k=1}^L\psi_j^{(n^{(k)}+1)},
        \end{aligned}
    \end{equation}
    then determinant $\tau(n)=\textnormal{det}(m_{ij}^{(n)})$ satisfies

    \begin{equation}
        \begin{aligned}
            \notag
            &(D_{x_1}^2-2r_kD_{x_1}-D_{x_2})\tau(n^{(k)}+1)\cdot \tau(n)=0,\\
            &(D_{x_1}^2+2r_kD_{x_1}+D_{x_2})\tau(n^{(k)}-1)\cdot \tau(n)=0,\\
            &D_{x_1}D_{x_{-1}}\tau(n)\cdot \tau(n)=-2\sum\limits_{k=1}^L(\tau(n^{(k)}+1)\tau(n^{(k)}-1)-\tau(n)^2).
        \end{aligned}
    \end{equation}
\end{theorem}

\begin{proof}
    For determinant $\text{det}(a_{ij})_{1\leq i,j\leq N}$ we have
    \begin{equation}
        \begin{aligned}
            \notag
            &\partial_x \text{det}(a_{ij})=\sum\limits_{i,j=1}^N\Delta_{ij}\partial_x a_{ij},\\
            &\text{det}\left(\begin{matrix}
                a_{ij} & b_i \\
                c_j & d
            \end{matrix}\right)=-\sum\limits_{i,j=1}^N\Delta_{ij}b_ic_j+d \text{det}(a_{ij}),
        \end{aligned}
    \end{equation}
    Thus for determinant satisfying the properties of theorem,
    \begin{equation}
        \begin{aligned}
            \notag
            &\partial_{x_1}\tau(n)=\text{det}\left(\begin{matrix}
                m_{ij}^{(n)} & \phi_i^{(n)}\\
                -\psi_j^{(n)} & 0
            \end{matrix}\right),\\
            &\partial_{x_1}^2\tau(n)=\text{det}\left(\begin{matrix}
                m_{ij}^{(n)} & \partial_{x_1}\phi_i^{(n)}\\
                -\psi_j^{(n)} & 0
            \end{matrix}\right)+\text{det}\left(\begin{matrix}
                m_{ij}^{(n)} & \phi_i^{(n)}\\
                -\partial_{x_1}\psi_j^{(n)} & 0
            \end{matrix}\right),\\
            &\partial_{x_2}\tau(n)=\text{det}\left(\begin{matrix}
                m_{ij}^{(n)} & \partial_{x_1}\phi_i^{(n)}\\
                -\psi_j^{(n)} & 0
            \end{matrix}\right)-\text{det}\left(\begin{matrix}
                m_{ij}^{(n)} & \phi_i^{(n)}\\
                -\partial_{x_1}\psi_j^{(n)} & 0
            \end{matrix}\right),\\
            &\partial_{x_{-1}}\tau(n)=\sum\limits_{k=1}^L\text{det}\left(\begin{matrix}
                m_{ij}^{(n)} & \phi_i^{(n^{(k)}-1)}\\
                \psi_j^{(n^{(k)}+1)} & 0
            \end{matrix}\right),\\
        \end{aligned}
    \end{equation}
    \begin{equation}
        \begin{aligned}
            \notag
            &\partial_{x_1}\partial_{x_{-1}}\tau(n)=L\text{det}(m_{ij}^{(n)})+\sum\limits_{k=1}^L\text{det}\left(\begin{matrix}
                m_{ij}^{(n)} & \phi_i^{(n^{(k)}-1)} & \phi_i^{(n)}\\
                \psi_j^{(n^{(k)}+1)} & 0 & -1\\
                -\psi_j^{(n)} &-1 & 0
            \end{matrix}\right),\\
            &\tau(n^{(k)}+1)=\text{det}\left(\begin{matrix}
                m_{ij}^{(n)} & \phi_i^{(n)}\\
                -\psi_j^{(n^{(k)}+1)} & 1
            \end{matrix}\right),\quad
            \tau(n^{(k)}-1)=\text{det}\left(\begin{matrix}
                m_{ij}^{(n)} & \phi_i^{(n^{(k)}-1)}\\
                \psi_j^{(n)} & 1
            \end{matrix}\right),\\
            &(\partial_{x_1}-r_k)\tau(n^{(k)}+1)=\text{det}\left(\begin{matrix}
                m_{ij}^{(n)} & \partial_{x_1}\phi_i^{(n)}\\
                -\psi_j^{(n^{(k)}+1)} & -r_k
            \end{matrix}\right),\\
            &(\partial_{x_1}-r_k)^2\tau(n^{(k)}+1)\\
            &=\text{det}\left(\begin{matrix}
                m_{ij}^{(n)} & \partial_{x_1}^2\phi_i^{(n)}\\
                -\psi_j^{(n^{(k)}+1)} & r_k^2
            \end{matrix}\right)+\text{det}\left(\begin{matrix}
                m_{ij}^{(n)} & \phi_i^{(n)} & \partial_{x_1}\phi_i^{(n)}\\
                -\psi_j^{(n)} & 0 & 0\\
                -\psi_j^{(n^{(k)}+1)} & 1 & -r_k
            \end{matrix}\right),\\
            &(\partial_{x_2}+r_k^2)\tau(n^{(k)}+1)\\
            &=\text{det}\left(\begin{matrix}
                m_{ij}^{(n)} & \partial_{x_1}^2\phi_i^{(n)}\\
                -\psi_j^{(n^{(k)}+1)} & r_k^2
            \end{matrix}\right)-\text{det}\left(\begin{matrix}
                m_{ij}^{(n)} & \phi_i^{(n)} & \partial_{x_1}\phi_i^{(n)}\\
                -\psi_j^{(n)} & 0 & 0\\
                -\psi_j^{(n^{(k)}+1)} & 1 & -r_k
            \end{matrix}\right).
        \end{aligned}
    \end{equation}

    Substituting to
    \begin{equation}
        \begin{aligned}
            \notag
            &(D_{x_1}^2-2r_kD_{x_1}-D_{x_2})\tau(n^{(k)}+1)\cdot \tau(n)=0,\\
            &D_{x_1}D_{x_{-1}}\tau(n)\cdot \tau(n)=-2\sum\limits_{k=1}^L(\tau(n^{(k)}+1)\tau(n^{(k)}-1)-\tau(n)^2),\\
        \end{aligned}
    \end{equation}
    noting that the consequence are exactly the Jacobi identity

    \begin{equation}
        \begin{aligned}
            \notag
            \text{det}\left(\begin{matrix}
                a_{ij} & b_i &c_i \\
                d_j & e & f \\
                g_j & h & k
            \end{matrix}\right)\text{det}(a_{ij})&=\text{det}\left(\begin{matrix}
                a_{ij} & c_i \\
                g_j & k
            \end{matrix}\right)\text{det}\left(\begin{matrix}
                a_{ij} & b_i \\
                d_j & e
            \end{matrix}\right)\\
            &-\text{det}\left(\begin{matrix}
                a_{ij} & b_i \\
                g_j & h
            \end{matrix}\right)\text{det}\left(\begin{matrix}
                a_{ij} & c_i \\
                d_j & f
            \end{matrix}\right),\\
        \end{aligned}
    \end{equation}
    So the first and the third equalities in theorem hold.

    Note that the equation
    $$(D_{x_1}^2-2r_kD_{x_1}-D_{x_2})\tau(n^{(k)}+1)\cdot \tau(n)=0$$
	is equivalent to the equation
    $$(D_{x_1}^2+2r_kD_{x_1}+D_{x_2})\tau(n^{(k)}-1)\cdot \tau(n)=0,$$
    therefore, the second equation in Theorem $\ref{LNgram}$ also holds.
\end{proof}

If we let
$$x_1=x,\quad x_{-1}=y,\quad x_2=t,$$
then Theorem $\ref{LNgram}$ provides a solution to the equations $(\ref{nengsuanshuangxianxing})$ given by
$$F=\tau(n),\quad G_k^+=\tau(n^{(k)}+1),\quad G_k^-=\tau(n^{(k)}-1),\quad  (n)=(n^{(1)},n^{(2)},\cdots ,n^{(L)}).$$

Let $\tilde{M}{ij}^{(n)}$, $\tilde{\phi}i^{(n)}$, $\tilde{\psi}j^{(n)}$ be functions of $x_1$, $x{-1}$, $x_2$ $(1\leq i,j\leq N)$,
where $(n)=(n^{(1)},n^{(2)},\cdots ,n^{(L)})$, and are specifically constructed as
\begin{equation}\label{NMPP}
    \begin{aligned}
        &\tilde{M}_{ij}^{(n)}=A_iB_j\frac{1}{p+q}\prod\limits_{k=1}^L(-\frac{p+r_k}{q-r_k})^{n^{(k)}}e^{\tilde{\xi}+\tilde{\eta}},\\
        &\tilde{\phi}_i^{(n)}=A_i\prod\limits_{k=1}^L(p+r_k)^{n^{(k)}}e^{\tilde{\xi}},\\
        &\tilde{\psi}_j^{(n)}=B_j\prod\limits_{k=1}^L(-q+r_k)^{-n^{(k)}}e^{\tilde{\eta}},
    \end{aligned}
\end{equation}
where
\begin{equation}
    \begin{aligned}
        \notag
        \tilde{\xi}=\sum\limits_{k=1}^L\frac{1}{p+r_k}x_{-1}+px_1+p^2x_2,\\
        \tilde{\eta}=\sum\limits_{k=1}^L\frac{1}{q-r_k}x_{-1}+qx_1-q^2x_2,\\
    \end{aligned}
\end{equation}
and $A_i$, $B_j$ are differential operators with respect to $p$ and $q$, respectively,
$$A_i=\sum\limits_{k=0}^{i}a_{ik}\partial_{p}^{i-k},\quad B_j=\sum\limits_{l=0}^{j}b_{jl}\partial_{q}^{j-l},$$
where $a_{ik}$, $b_{jl}$ are real parameters of the operators.

\begin{lemma}\label{lnfit}
    The functions $(\ref{NMPP})$ satisfy the requirements of Theorem $\ref{LNgram}$, and thus provide a solution to 
    equation $(\ref{nengsuanshuangxianxing})$.
\end{lemma}

Lemma $\ref{lnfit}$ can be directly valided.

\section{Weak reduction condition}

To continue solving equation $(\ref{LNRD})$, a condition reducing the given bilinear equation needs to be consider. For the function 
$\tilde{M}_{ij}^{(0)}$ defined by $(\ref{NMPP})$ with parameters $p$, $q$, $x_{-1}$, $x_1$, $x_2$, if we can determine the parameters 
$K=K(r_1,\cdots ,r_L)$ through $r_1$, $\cdots $, $r_L$, such that there exist $p$, $q$ satisfying
\begin{equation}\label{WYH}
    \tag{*}
    \begin{aligned}
        (K\partial_{x_1}+\partial_{x_{-1}})\tilde{M}_{ij}^{(0)}=C\tilde{M}_{ij}^{(0)}
        +\sum\limits_kC_k\tilde{M}_{i_kj}^{(0)}+\sum\limits_kC_k^{\prime}\tilde{M}_{ij_k}^{(0)},
    \end{aligned}
\end{equation}
where $C$, $C_k$, $C_k^{\prime}$ are constants, $i_k$, $j_k$ are integer sequences satisfying $i_k<i$, $j_k<j$, then we have
\begin{equation}
    \begin{aligned}
        \notag
        &(K\partial_{x_1}+\partial_{x_{-1}})\text{det}(\tilde{M}_{ij}^{(0)})\\
        =&\sum\limits_{i,j}\Delta_{i,j}(K\partial_{x_1}+\partial_{x_{-1}})\tilde{M}_{ij}^{(0)}\\
        \notag
        =&C\sum\limits_{i,j}\Delta_{i,j}\tilde{M}_{ij}^{(0)}+\sum\limits_{k}C_k
        \sum\limits_{i,j}\Delta_{i,j}\tilde{M}_{i_kj}^{(0)}+\sum\limits_{k}C_k^{\prime}
        \sum\limits_{i,j}\Delta_{i,j}\tilde{M}_{ij_k}^{(0)}\\
        =&C\text{det}(\tilde{M}_{ij}^{(0)}).
    \end{aligned}
\end{equation}

where $C$, $C_k$, $C_k^{\prime}$ are constants, $i_k$, $j_k$ are integer sequences satisfying $i_k<i$, $j_k<j$, then we have
$$D_{x_{-1}}D_{x_1}F\cdot F=-KD_{x_1}^2F\cdot F,$$
with $F=\text{det}(\tilde{M}_{ij}^{(0)})$.
Then we can get a solution to the bilinear equation
\begin{equation}\label{Kshuangxianxing}
    \begin{aligned}
        &(D_x^2-2r_kD_x-D_t)G_k^+\cdot F=0,\\
        &(D_x^2+2r_kD_x+D_t)G_k^-\cdot F=0,\\
        &KD_x^2F\cdot F=2\sum\limits_{k=1}^L (G_k^+G_k^--F^2).
    \end{aligned}
\end{equation}

This condition is slightly weaker than the KP reduction and is therefore referred to as a weak reduction condition. 
In the following we always assume that the reduction condition holds.

\section{Rational solutions to the RNLS equation}

This section presents the main theorem, which provides a general rational solution for the nonlocal 
discrete RD system $(\ref{LNRD})$ under weak reduction conditions.

\begin{theorem}\label{lnthm}
    For parameters $r_1,\cdots ,r_L$, if there exists $K$ such that condition $(*)$ holds, then the discrete nonlocal 
    RD system $(\ref{LNRD})$ has a rational solution given by
    \begin{equation}
        \begin{aligned}
            \notag
            &e^+=\frac{e^{\frac{2L}{K}t-r_kx+r_k^2t}}{\sqrt{|K|}}\frac{\tau(0^{(k)}+1)}{\tau(0)},\\
            &e^-=-\frac{e^{-\frac{2L}{K}t+r_kx-r_k^2t}}{\sqrt{|K|}}\frac{\tau(0^{(k)}-1)}{\tau(0)},\\
        \end{aligned}
    \end{equation}

    where $\tau(n)=\textnormal{det}(m_{ij}^{(n)})$ is an $N$-th order determinant, $(n)$ is an $L$-element integer array, 
    \begin{equation}
        \begin{aligned}
            \notag
            m_{ij}^{(n)}=&[\prod\limits_{k=1}^L(-\frac{p+r_k}{q-r_k})^{n^{(k)}}]
            [\sum\limits_{k=0}^{i}a_{ik}(\sum\limits_{l=1}^{L}\frac{n^{(l)}}{p+r_l}+\xi^{\prime}+\partial_{p})^{i-k}]\\
            &[\sum\limits_{k=0}^{j}b_{jk}(-\sum\limits_{l=1}^{L}\frac{n^{(l)}}{q-r_l}+\eta^{\prime}+\partial_{q})^{j-l}]
            \frac{1}{p+q}|_{(*)},
        \end{aligned}
    \end{equation}
    $$\xi^{\prime}=x+2pt,\quad \eta^{\prime}=x-2qt.$$
\end{theorem}

\begin{proof}
    Take $x_{-1}=0$, $x_1=x$, and $x_2=t$ in equation $(\ref{NMPP})$ to obtain
    $$M_{ij}^{(n)}=A_iB_j\frac{1}{p+q}\prod\limits_{k=1}^L(-\frac{p+r_k}{q-r_k})^{n^{(k)}}e^{\xi+\eta},$$
    where
    \begin{equation}
        \begin{aligned}
            \notag
            \xi=px+p^2t,\quad \eta=qx-q^2t.
        \end{aligned}
    \end{equation}

    define the determinant
    $$\tau(n)=\text{det}(M_{ij}^{(n)}),$$
    Given the weak reduction condition $(*)$, for any array $(n)$, the system $(\ref{Kshuangxianxing})$ of equations can be solved by
    $$(F,\ G_k^+,\ G_k^-)=(\tau(n),\ \tau(n^{(k)+1}),\ \tau(n^{(k)-1})).$$
    
    If the weak reduction condition $(*)$ has $K>0$, we can transform the system $(\ref{LNRD})$ as follows
    $$e^+_k\mapsto \sqrt{K}e^+_ke^{-\frac{2Lt}{K}},\quad e^-_k\mapsto -\sqrt{K}e^-_ke^{\frac{2Lt}{K}},$$
    which yields
    \begin{equation}\label{KLNRD}
        \begin{aligned}
            -&e^+_{k,t}+e^+_{k,xx}=\frac{2}{K}(h+L)e^+_k,\\
            &e^-_{k,t}+e^-_{k,xx}=\frac{2}{K}(h+L)e^-_k,\\
            &h=-\sum\limits_{k=1}^Le^+_ke^-_k.
        \end{aligned}
    \end{equation}
    By using the rational transformation
    $$e^{+}_k=e^{-r_kx+r_k^2t}\frac{G^{+}_k}{F},\quad e^{-}_k=e^{r_kx-r_k^2t}\frac{G^{-}_k}{F},$$
    the system $(\ref{Kshuangxianxing})$ can be obtained, and thus if the equation $(\ref{Kshuangxianxing})$ has a 
    solution $(F,\ G_k^+,\ G_k^-)$, then the equation $(\ref{LNRD})$ has a solution
    \begin{equation}\label{LNbianhuan}
        \begin{aligned}
            e^+=\frac{e^{\frac{2L}{K}t-r_kx+r_k^2t}}{\sqrt{K}}\frac{G^+_k}{F},\quad
            e^-=-\frac{e^{-\frac{2L}{K}t+r_kx-r_k^2t}}{\sqrt{K}}\frac{G^-_k}{F}.
        \end{aligned}
    \end{equation}
    For the case where $K<0$, a similar transformation can be performed on $-K$, and the equation $(\ref{LNRD})$ has the solution
    \begin{equation}
        \begin{aligned}
            \notag
            e^+=\frac{e^{\frac{2L}{K}t-r_kx+r_k^2t}}{\sqrt{-K}}\frac{G^+_k}{F},\quad
            e^-=\frac{e^{-\frac{2L}{K}t+r_kx-r_k^2t}}{\sqrt{-K}}\frac{G^-_k}{F}.
        \end{aligned}
    \end{equation}
    
    By direct calculation, the function $(\ref{NMPP})$ satisfies
    \begin{equation}
        \begin{aligned}
            \notag
            \tilde{M}_{ij}^{(n)}=&[\prod\limits_{k=1}^L(-\frac{p+r_k}{q-r_k})^{n^{(k)}}]e^{\tilde{\xi}+\tilde{\eta}}
            [\sum\limits_{k=0}^{i}a_{ik}(\sum\limits_{l=1}^{L}\frac{n^{(l)}}{p+r_l}+\tilde{\tilde{\xi}}+\partial_{p})^{i-k}]\\
            &[\sum\limits_{k=0}^{j}b_{jk}(-\sum\limits_{l=1}^{L}\frac{n^{(l)}}{q-r_l}+\tilde{\tilde{\eta}}+\partial_{q})^{j-l}]\frac{1}{p+q},
        \end{aligned}
    \end{equation}
    where
    \begin{equation}
        \begin{aligned}
            \notag
            \tilde{\tilde{\xi}}=-\sum\limits_{k=1}^L\frac{1}{(p+r_k)^2}x_{-1}+x_1+2px_2,\\
            \tilde{\tilde{\eta}}=-\sum\limits_{k=1}^L\frac{1}{(q-r_k)^2}x_{-1}+x_1-2qx_2.\\
        \end{aligned}
    \end{equation}

    Therefore, if there exists a $K$ that satisfies the weak reduction condition $(*)$, then the equation $(\ref{Kshuangxianxing})$ has a solution
    $$(F,\ G_k^+,\ G_k^-)=(\tau(n),\ \tau(n^{(k)+1}),\ \tau(n^{(k)-1})),$$
    where $\tau(n)=\text{det}(M_{ij}^{(n)})$, and
    \begin{equation}
        \begin{aligned}
            \notag
            M_{ij}^{(n)}=&[\prod\limits_{k=1}^L(-\frac{p+r_k}{q-r_k})^{n^{(k)}}]e^{\xi+\eta}
            [\sum\limits_{k=0}^{i}a_{ik}(\sum\limits_{l=1}^{L}\frac{n^{(l)}}{p+r_l}+\xi^{\prime}+\partial_{p})^{i-k}]\\
            &[\sum\limits_{k=0}^{j}b_{jk}(-\sum\limits_{l=1}^{L}\frac{n^{(l)}}{q-r_l}+\eta^{\prime}+\partial_{q})^{j-l}]\frac{1}{p+q},
        \end{aligned}
    \end{equation}
    with
    $$\xi=px+p^2t,\quad \eta=qx-q^2t,$$
    $$\xi^{\prime}=x+2pt,\quad \eta^{\prime}=x-2qt.$$
    Note that the exponential term of the denominator of the numerator of the rational function can be eliminated, 
    which completes the proof of the theorem.

\end{proof}

The result of Theorem $\ref{lnthm}$ depends on the weak reduction condition $(*)$, but the condition itself 
may not hold. We discuss the first-order rational solution of the discrete nonlocal RD system $(\ref{LNRD})$, 
and the general high-order rational solutions of the RNLS equation $(\ref{RNLS})$, 
to show that the condition $(*)$ is non-trivial.

\section{The case $L=1$ for arbitrary $N$}

The following lemma shows that the reduction condition holds for $L=1$, and hence the function $(\ref{NMPP})$ 
gives a solution to the RNLS equation $(\ref{RNLS})$ \cite{ref22}.

\begin{lemma}\label{yuehua}
    When $L=1$, for the function $(\ref{NMPP})$, with $p,\ q=1$ we have
    $$\partial_{x_{-1}}\textnormal{det}(\tilde{M}_{2i-1,2j-1}^{(n)})
    +\partial_{x_1}\textnormal{det}(\tilde{M}_{2i-1,2j-1}^{(n)})
    =C\textnormal{det}(\tilde{M}_{2i-1,2j-1}^{(n)}),$$

    Thus, for any fixed value of $y$, $\tau_n=\textnormal{det}(\tilde{M}_{ij}^{(n)})$ gives a solution 
    $$(F,G^+,G^-)=(\tau_n,\tau_{n+1},\tau_{n-1})|_{y=y_0,p=q=1}.$$
\end{lemma}

When $L=1$, we can simplify the solution by Schur polynomials defined by
\begin{definition}\label{schur}
	For indeterminates $x_1,x_2,\cdots$, $S_k$ is a $k$-degree polynomial with respect to $x_1,\cdots,x_k$ (called the Schur polynomial), 
    where $k\geq 0$, defined as
	$$\sum\limits_{k=0}^{\infty}S_k\lambda^k=e^{\sum\limits_{k=1}^{\infty}x_k\lambda^k},$$
    where the left side is a formal power series of the right side.
\end{definition}

\begin{lemma}\label{shuajian}
    There exist polynomials $P_i^{(n)}$, $Q_j^{(n)}$ of degree $i$ and $j$, respectively, such that
    $$A_ip^ne^{\tilde{\xi}}=P_i^{(n)}p^ne^{\tilde{\xi}},$$
    $$B_j(-q)^{-n}e^{\tilde{\eta}}=Q_j^{(n)}(-q)^{-n}e^{\tilde{\eta}}.$$
\end{lemma}

\begin{proof}
    Note that for any differentiable function $F$, we have the differential equation
    $$e^{\lambda p\partial_p}F(p)=F(e^{\lambda}p),$$
    Thus
    \begin{equation}
        \begin{aligned}
            \notag
            \frac{1}{p^ne^{\xi}}e^{\lambda p\partial_p}p^ne^{\xi}
            =&\frac{1}{p^ne^{\sum\limits p^{\nu}x_{\nu}}}e^{\lambda p\partial_p}p^ne^{\sum\limits p^{\nu}x_{\nu}}\\
            =&\frac{1}{p^ne^{\sum\limits p^{\nu}x_{\nu}}}(e^{\lambda}p)^ne^{\sum\limits (e^{\lambda}p)^{\nu}x_{\nu}}\\
            =&e^{\lambda n+\sum\limits_{\nu}\sum\limits_{k=1}^{\infty}\frac{(\nu\lambda)^k}{k!}p^{\nu}x_{\nu}}\\
            =&e^{\lambda n+\sum\limits_{k=1}^{\infty}\frac{\lambda^k}{k!}\sum\limits_{\nu}\nu^kp^{\nu}x_{\nu}},
        \end{aligned}
    \end{equation}
    thus, by the definition of Schur polynomials $S_k$, we have
    $$\frac{1}{p^ne^{\xi}}e^{\lambda p\partial_p}p^ne^{\xi}=\sum\limits_{k=0}^{\infty}\lambda^kS_k(\xi^{(n)}(p)),$$
    where
    $$\xi^{(n)}(p)=(\xi^{(n)}_1(p)+n,\ \xi^{(n)}_2(p),\cdots ,\xi^{(n)}_k(p),\cdots ),$$
    $$\xi^{(n)}_k(p)=\frac{(-1)^k\frac{1}{p}x_{-1}+px_1+2^kp^2x_2}{k!},$$
    thus by pending coefficient method, we obtain
    $$(p\partial p)^kp^ne^{\xi}=k!p^ne^{\xi}S_k(\xi^{(n)}(p)),$$
    $$P_i^{(n)}=\sum\limits_{k=0}^{i}a_{k}S_{i-k}(\xi^{(n)}(p)).$$
    
    Similarly
    $$Q_j^{(n)}=\sum\limits_{l=0}^{j}b_{l}S_{j-l}(\eta^{(n)}(q)),$$
    where
    $$\eta^{(n)}(q)=(\eta^{(n)}_1(q)-n,\ \eta^{(n)}_2(q),\cdots ,\eta^{(n)}_k(q),\cdots ),$$
    $$\eta^{(n)}_k(q)=\frac{(-1)^k\frac{1}{q}x_{-1}+qx_1-2^kq^2x_2}{k!}.$$
    
    It can be verified that the above $P_i^{(n)}$, $Q_j^{(n)}$ satisfy the requirements of the Lemma. 
\end{proof}

the function $(\ref{NMPP})$ satisfies
$$\partial_{x_1}\tilde{M}_{ij}^{(n)}=\tilde{\phi}_i^{(n)}\tilde{\psi}_j^{(n)},$$
Integrate by parts, we obtain
\begin{equation}
    \begin{aligned}
        \notag
        \tilde{M}_{ij}^{(n)}=&\int\tilde{\phi}_i^{(n)}\tilde{\psi}_j^{(n)}dx_1\\
        =&c_{ij}+(-\frac{p}{q})^ne^{\tilde{\xi}+\tilde{\eta}}\sum\limits_{k=0}^{i+j}
        \frac{(-1)^{k}}{(p+q)^{k+1}}\partial_{x_1}^{k}P_i^{(n)}Q_j^{(n)},
    \end{aligned}
\end{equation}

Take $x_{-1}=0$, for a rational transformation, the exponential factors in the numerator and denominator 
can be canceled simultaneously, so that $(\ref{LNRD})$ has a solution
$$e^+=\frac{\text{det}(m_{2i-1,2j-1}^{(n+1)})}{\text{det}(m_{2i-1,2j-1}^{(n)})},\quad
e^-=\frac{\text{det}(m_{2i-1,2j-1}^{(n-1)})}{\text{det}(m_{2i-1,2j-1}^{(n)})},$$
where
$$m_{ij}^{(n)}=(-\frac{p}{q})^n\sum\limits_{k=0}^{i+j}
\frac{(-1)^{k}}{(p+q)^{k+1}}\partial_{x_1}^{k}P_i^{(n)}Q_j^{(n)}|_{p=1,q=1}.$$

In this case, the weak reduction condition holds with $K=1$, and thus is the conventional KP reduction method.
That's why it is called the weak reduction condition.

$\bullet$one-order rational solutions.

When $N=1$, we have
$$\tau_n=m_{11}^{(n)},$$
Using the definition of $m_{11}^{(n)}$, we obtain
\begin{equation}
    \begin{aligned}
        \notag
        m_{11}^{(n)}=&\frac{(a_0(x+2t+n)+a_1)(b_0(x-2t-n)+b_1)}{2}\\
        &-\frac{a_0(b_0(x-2t-n)+b1)}{4}-\frac{b_0(a_0(x+2t+n)+a_1)}{4}+\frac{a_0b_0}{4},
    \end{aligned}
\end{equation}
Taking $a_0=b_0=1$, $a_1=b_1=0$, we have
\begin{equation}
    \begin{aligned}
        \notag
        \tau_n=\frac{(x+2t+n)(x-2t-n)}{2}-\frac{(x-2t-n)}{4}-\frac{(x+2t+n)}{4}+\frac{1}{4},
    \end{aligned}
\end{equation}
Thus, the RD system has the solution
\begin{equation}
    \begin{aligned}
        \notag
        e^+=&\frac{\tau_1}{\tau_0}
        =1-\frac{4t+1}{x^2-x-4t^2+\frac{1}{2}},\\
        e^-=&\frac{\tau_{-1}}{\tau_0}
        =1+\frac{4t-1}{x^2-x-4t^2+\frac{1}{2}},\\
    \end{aligned}
\end{equation}
which represents an algebraic 2-soliton. At this point, in the following two planar regions,

(1)Region $\{(x,t)|x>-2t+\frac{3}{2},\ x>2t+\frac{3}{2}\}$.

(2)Region $\{(x,t)|-2t+2<x<2t+1\}$.

The RNLS equation $(\ref{RNLS})$ has the solution
\begin{equation}
    \begin{aligned}
        \notag
        &|\Psi|^2=1-\frac{2}{x^2-x-4t^2+\frac{1}{2}}-\frac{16t^2-1}{(x^2-x-4t^2+\frac{1}{2})^2},\\
        &\text{arg}\Psi=\frac{1}{4}\text{ln}(1+\frac{8t}{x^2-x-4t^2-4t+\frac{1}{2}})-t.
    \end{aligned}
\end{equation}

$\bullet$ 2-order rational solution.

When $N=2$, we have
$$\tau_n=m_{11}^{(n)}m_{33}^{(n)}-m_{13}^{(n)}m_{31}^{(n)},$$
$$e^+=\frac{\tau_1}{\tau_0},\quad e^-=\frac{\tau_{-1}}{\tau_0}.$$

In this case, we have two free parameters $a_3$ and $b_3$. Depending on the values of these two parameters, 
the equation has three types of solutions, which are:

1.When $a_3=b_3=0$, we have
\begin{equation}
    \begin{aligned}
        \notag
        \tau_0=&\frac{x^6}{5184}-\frac{x^5}{1728}-\frac{(8t^2-3)x^4}{3456}+\frac{(16t^2-3)x^3}{3456}+\frac{(32t^4+3)x^2}{3456}\\
        &-\frac{(64t^4+24t^2+3)x}{6912}-\frac{t^6}{81}+\frac{5t^4}{216}-\frac{t^2}{288}+\frac{1}{9216},\\
        \tau_1=&\tau_0-\frac{t^5}{27}-\frac{5t^4}{108}+\frac{(2x^2-2x+1)t^3}{108}+\frac{(x^2-x+1)t^2}{72}\\
        &-\frac{(2x^4-4x^3+3x-3)t}{864}-\frac{x^4}{1728}+\frac{x^3}{864}-\frac{x^2}{576}+\frac{x}{1152},\\
        \tau_{-1}=&\tau_0+\frac{t^5}{27}-\frac{5t^4}{108}-\frac{(2x^2-2x+1)t^3}{108}+\frac{(x^2-x+1)t^2}{72}\\
        &+\frac{(2x^4-4x^3+3x-3)t}{864}-\frac{x^4}{1728}+\frac{x^3}{864}-\frac{x^2}{576}+\frac{x}{1152}.\\
    \end{aligned}
\end{equation}

In this case, the RD system has solutions in the two regions below on the plane, and the equation $(\ref{RNLS})$ has the solutions, 
where $\tau_0$, $\tau_1$, and $\tau_{-1}$ are given as above.

(1)Region $\{(x,t)|x>-2t+3,\ x>2t+3\}$;

(2)Region $\{(x,t)|-2t+\frac{9}{2}<x<2t-\frac{7}{2}\}$.

2.When $a_3b_3>0$, we take $a_3=b_3=1$, and we have
\begin{equation}
    \begin{aligned}
        \notag
        \tau_0=&\frac{x^6}{5184}-\frac{x^5}{1728}-\frac{(8t^2-3)x^4}{3456}+\frac{(16t^2-27)x^3}{3456}+\frac{(32t^4+39)x^2}{3456}\\
        &-\frac{(64t^4+600t^2+3)x}{6912}-\frac{t^6}{81}+\frac{5t^4}{216}+\frac{11t^2}{288}+\frac{577}{9216},\\
        \tau_1=&\tau_0-\frac{t^5}{27}-\frac{5t^4}{108}+\frac{(2x^2-2x+1)t^3}{108}+\frac{(x^2-x+1)t^2}{72}\\
        &-\frac{(2x^4-4x^3+75x-39)t}{864}-\frac{x^4}{1728}+\frac{x^3}{864}-\frac{x^2}{576}-\frac{23x}{1152}+\frac{1}{96},\\
        \tau_{-1}=&\tau_0+\frac{t^5}{27}-\frac{5t^4}{108}-\frac{(2x^2-2x+1)t^3}{108}+\frac{(x^2-x+1)t^2}{72}\\
        &+\frac{(2x^4-4x^3+75x-39)t}{864}-\frac{x^4}{1728}+\frac{x^3}{864}-\frac{x^2}{576}-\frac{23x}{1152}+\frac{1}{96}.\\
    \end{aligned}
\end{equation}

In this case, the RD system has solutions in the two regions below on the plane, and the equation $(\ref{RNLS})$ has the solutions, 
where $\tau_0$, $\tau_1$, and $\tau_{-1}$ are given as above.

(1)Region $\{(x,t)|x>-2t+5,\ x>2t+5\}$;

(2)Region $\{(x,t)|-2t+5<x<2t+1\}$.

3.When $a_3b_3<0$, we take $a_3=1$, $b_3=-1$, and we have
\begin{equation}
    \begin{aligned}
        \notag
        \tau_0=&\frac{x^6}{5184}-\frac{x^5}{1728}-\frac{(8t^2-3)x^4}{3456}+\frac{(16t^2-3)x^3}{3456}+\frac{(32t^4+144t+3)x^2}{3456}\\
        &-\frac{(64t^4+24t^2+288t+3)x}{6912}-\frac{t^6}{81}+\frac{5t^4}{216}+\frac{t^3}{18}-\frac{t^2}{288}-\frac{t}{48}-\frac{575}{9216},\\
        \tau_1=&\tau_0-\frac{t^5}{27}-\frac{5t^4}{108}+\frac{(2x^2-2x+1)t^3}{108}+\frac{(x^2-x+7)t^2}{72}\\
        &-\frac{(2x^4-4x^3+3x-39)t}{864}-\frac{x^4}{1728}+\frac{x^3}{864}+\frac{11x^2}{576}-\frac{23x}{1152}+\frac{1}{96},\\
        \tau_{-1}=&\tau_0+\frac{t^5}{27}-\frac{5t^4}{108}-\frac{(2x^2-2x+1)t^3}{108}+\frac{(x^2-x+7)t^2}{72}\\
        &+\frac{(2x^4-4x^3+3x-39)t}{864}-\frac{x^4}{1728}+\frac{x^3}{864}-\frac{13x^2}{576}+\frac{25x}{1152}-\frac{1}{96}.\\
    \end{aligned}
\end{equation}

In this case, the RD system has solutions in the two regions below on the plane, and the equation $(\ref{RNLS})$ has the solutions,
where $\tau_0$, $\tau_1$, and $\tau_{-1}$ are given as above.

(1)Region $\{(x,t)|x>-2t+5,\ x>2t-1\}$;

(2)Region $\{(x,t)|-2t+\frac{11}{2}<x<2t-\frac{9}{2}\}$.

\section{The case $N=1$ for arbitrary $L$}

Consider equation $(\ref{LNRD})$, and let
\begin{equation}
    \begin{aligned}
        \notag
        u=(\sum\limits_{k=1}^L\frac{1}{p+r_k}+\sum\limits_{k=1}^L\frac{1}{q-r_k})y+(p+q)x+(p^2-q^2)t,
    \end{aligned}
\end{equation}
By Lemma $\ref{lnfit}$, the bilinear equation $(\ref{nengsuanshuangxianxing})$ has a solution
$$F=\partial_p\partial_q\frac{1}{p+q}e^u.$$

Note that
$$u_{pq}=0,$$
thus for the function $F$ and its derivatives, we have
\begin{equation}
    \begin{aligned}
        \notag
        \partial_xF=&e^uu_pu_q,\\
        \partial_yF=&(\sum\limits_{k=1}^L\frac{1}{(p+r_k)^2(q-r_k)^2})e^u-(\sum\limits_{k=1}^L\frac{1}{(p+r_k)(q-r_k)^2})e^uu_p\\
        &-(\sum\limits_{k=1}^L\frac{1}{(p+r_k)^2(q-r_k)})e^uu_q+(\sum\limits_{k=1}^L\frac{1}{(p+r_k)(q-r_k)})e^uu_pu_q,\\
        F=&\frac{2}{(p+q)^3}e^u-\frac{1}{(p+q)^2}e^uu_p-\frac{1}{(p+q)^2}e^uu_q+\frac{1}{p+q}e^uu_pu_q.
    \end{aligned}
\end{equation}

It can be verified that when $p,\ q,\ r_k\ (1\leq k\leq L)$ satisfy the relation
\begin{equation}\label{pq}
    \begin{aligned}
        \sum\limits_{k=1}^L[\prod\limits_{l\neq k}(p+r_l)^2(q-r_l)^2](p-q+2r_k)=0
    \end{aligned}
\end{equation}
$F$ and $F_y$ have proportional coefficients with respect to $e^u$, $e^uu_p$, and $e^uu_q$. 
Thus, there exists a constant $K$ satisfying the weak reducibility condition $(*)$, where
\begin{equation}\label{K}
    \begin{aligned}
        K=\frac{1}{2}[\sum\limits_{k=1}^L\frac{1}{(p+r_k)^2}+\sum\limits_{k=1}^L\frac{1}{(q-r_k)^2}]>0.
    \end{aligned}
\end{equation}

The expression of $K$ given by $(\ref{K})$ satisfies the weak reduction condition $(*)$, therefore by Theorem $\ref{lnthm}$, 
the nonlocal RD system $(\ref{LNRD})$ has a solution given by
\begin{equation}\label{L1}
    \begin{aligned}
        e^+_k=\frac{e^{\frac{2Lt}{K}-r_kx+r_k^2t}}{\sqrt{K}}\frac{G^+_k}{F},
        \quad e^-_k=-\frac{e^{-\frac{2Lt}{K}+r_kx-r_k^2t}}{\sqrt{K}}\frac{G^-_k}{F},
    \end{aligned}
\end{equation}
where $p,\ q$ satisfy the relation $(\ref{pq})$, $K$ is given by $(\ref{K})$, and the functions $F,\ G^+_k,\ G^-_k$ are defined by
\begin{equation}
    \begin{aligned}
        \notag
        &F=H+\frac{2}{(p+q)^2},\\
        &G^+_k=-\frac{p+r_k}{q-r_k}H+\frac{R_k}{(q-r_k)^2}+\frac{2}{(p+q)^2},\\
        &G^-_k=-\frac{q-r_k}{p+r_k}H-\frac{R_k}{(p+r_k)^2}+\frac{2}{(p+q)^2},\\
    \end{aligned}
\end{equation}
with
\begin{equation}
    \begin{aligned}
        \notag
        &H=(x+2pt)(x-2qt)-\frac{2x+2(p-q)t}{p+q},\\
        &R_k=(p-q+2r_k)x+2(p^2+q^2+r_k(p-q))t.\\
    \end{aligned}
\end{equation}

Note that if $q$ is fixed and not equal to $r_k\ (1\leq k\leq L)$, then the relation $(\ref{pq})$ is a $2L-1$ degree polynomial 
equation with respect to $p$, and $-r_k\ (1\leq k\leq L)$ is not a root of the equation. Therefore, there always exist $p,q$ such that
$$p\neq -r_k,\quad q\neq r_k,\quad k=1,2,\cdots ,L,$$
and $p,\ q$ satisfy the relation $(\ref{pq})$. Thus, we have the following theorem.

\begin{theorem}
    Theorem $\ref{lnthm}$ gives a first-order rational solution $(\ref{L1})$ for the discrete nonlocal RD system $(\ref{LNRD})$.
\end{theorem}

For nonlocal RNLS equation, there is a solution that satisfies
$$\int_0^1|\Psi|^2d\xi=\frac{\sum\limits_{k=1}^LG^+_kG^-_k}{KF^2},$$
where $F,\ G^+_k,\ G^-_k$ are defined by $(\ref{L1})$.

For example, when $L=2$, $r_1=0$, $r_2=2$, $q=1$, we have $p=-1-\sqrt{3},\ K=2.$

\section{Bäcklund transformation of the RNLS equation}
We only study the local case in the remaining of this paper, i.e., take $L=1$.
By making the rational transformation of the equation $(\ref{LNRD})$ as
$$e^+=\frac{G^+}{F},\quad e^-=\frac{G^-}{F},$$
the bilinear RD system is obtained as
\begin{equation}\label{BRD}
    \begin{aligned}
        (D_x^2-D_t)G^+\cdot F=0,\\
        (D_x^2+D_t)G^-\cdot F=0,\\
        D_x^2F\cdot F=-2G^+G^-.
    \end{aligned}
\end{equation}

Taking two sets of solutions $(F,\ G^+,\ G^-)$ and $(F^{\prime},\ G^{+\prime},\ G^{-\prime})$ of the equation $(\ref{BRD})$, 
the two sets of solutions satisfy the equation
\begin{equation}\label{meihuajianback}
    \begin{aligned}
        &[(D_x^2-D_t)G^{+\prime}\cdot F^{\prime}]F^2-F^{\prime 2}[(D_x^2-D_t)G^+\cdot F]=0,\\
        &[(D_x^2+D_t)G^{-\prime}\cdot F^{\prime}]F^2-F^{\prime 2}[(D_x^2+D_t)G^-\cdot F]=0,\\
        &D_x^2F\cdot F=-2G^+G^-,\quad D_x^2F^{\prime}\cdot F^{\prime}=-2G^{+\prime}G^{-\prime}.
    \end{aligned}
\end{equation}

The following lemma provides the properties of the Hirota bilinear operator, which are the key to obtaining the Bäcklund transformation of 
the bilinear RD system $(\ref{BRD})$

\begin{lemma}\label{shuangxianxingprop}
    Hirota operator satisfies
        \begin{align}
            &(D_xa\cdot b)d^2-b^2(D_xc\cdot d)=D_x(a\cdot d+b\cdot c)bd-(ad+bc)D_xb\cdot d,\label{shuangxianxingpropa}\\
            &(D_x^2a\cdot b)d^2-b^2(D_x^2c\cdot d)=2D_x[D_x(a\cdot d+b\cdot c)\cdot bd]\notag\\
            &+[(ad-bc)D_x^2b\cdot d-D_x^2(a\cdot d-b\cdot c)bd]+[ab(D_x^2d\cdot d)-(D_x^2b\cdot b)cd],\label{shuangxianxingpropb}
        \end{align}
    where $a$, $b$, $c$, $d$ are functions about $x$.
\end{lemma}

Use this lemma we have
\begin{theorem}\label{backthm}
    Bilinear RD system$(\ref{BRD})$ has Bäcklund transformation
    \begin{equation}\label{Bäcklund}
        \begin{aligned}
            &D_t(G^{+\prime}\cdot F+F^{\prime}\cdot G^+)+(D_x^2-\lambda)(G^{+\prime}\cdot F- G^+\cdot F^{\prime})=0,\\
            &D_t(G^{-\prime}\cdot F+F^{\prime}\cdot G^-)-(D_x^2-\lambda)(G^{-\prime}\cdot F- G^-\cdot F^{\prime})=0,\\
            &(D_x^2+2\mu^2-\lambda)F^{\prime}\cdot F=-(G^{+\prime}G^-+G^{-\prime}G^+),\\
            &(D_t-2\mu D_x)F^{\prime}\cdot F=G^{+\prime}G^--G^{-\prime}G^+,\\
        \end{aligned}
    \end{equation}
    where $\lambda$, $\mu$ are arbitrary parameters, and $\mu$ satisfies
    \begin{equation}\label{xing}
        \begin{aligned}
            &D_x(G^{+\prime}\cdot F+F^{\prime}\cdot G^+)=\mu (G^{+\prime}F-F^{\prime} G^+),\\
            &D_x(G^{-\prime}\cdot F+F^{\prime}\cdot G^-)=-\mu (G^{-\prime}F-F^{\prime} G^-).
        \end{aligned}
    \end{equation}
\end{theorem}

If the equation $(\ref{Bäcklund})$ has a solution $(F,\ G^+,\ G^-,\ F^{\prime},\ G^{+\prime},\ G^{-\prime})$
and $(F^{\prime},\ G^{+\prime},\ G^{-\prime})$ is a solution of equation $(\ref{BRD})$, then $(F,\ G^+,\ G^-)$ gives
a solution of equation $(\ref{RNLS})$
\begin{equation}\label{psigpm}
    \begin{aligned}
        |\Psi|^2=-\frac{G^+G^-}{F^2},\quad 
        \text{arg}\Psi=\frac{1}{4}\text{ln}(-\frac{G^-}{G^+}).
    \end{aligned}
\end{equation}

\section{Soliton solutions of the RNLS eqaution}
This section gives the soliton solution of equation $(\ref{BRD})$ by using the Bäcklund transformation $(\ref{Bäcklund})$ provided by 
theorem $\ref{backthm}$. Using the Bäcklund transformation to solve equations requires a known solution of the original equation. 

$\bullet$ One-soliton solution

Take a particular solution of equation $(\ref{BRD})$ as
$$(F^{\prime},G^{+\prime},G^{-\prime})=(1,0,0),$$
substitute it into $(\ref{Bäcklund})$, and obtain
\begin{equation}\label{1jieguzi}
    \begin{aligned}
        &G^+_t+G^+_{xx}-\lambda G^+=0,\\
        &G^-_t-G^-_{xx}+\lambda G^-=0,\\
        &F_{xx}+(2\mu^2-\lambda)F=0,\\
        &F_t-2\mu F_x=0.\\
        &G^+_x=\mu G^+,\quad G^-_x=-\mu G^-
    \end{aligned}
\end{equation}
For equation $(\ref{1jieguzi})$, $F$, $G^{+}$, and $G^{-}$ can be solved separately.
To be concrete, we have:

1.when $\lambda>2\mu^2$, equation $(\ref{1jieguzi})$ has solutions
$$F(x,t)=C_1e^{\sqrt{\lambda-2\mu^2}(x+2\mu t)}+C_2e^{-\sqrt{\lambda-2\mu^2}(x+2\mu t)},$$
where $C_1$, $C_2$ are arbitrary constants.

2.For equation $(\ref{1jieguzi})$,
$$G^{+}(x,t)=Ce^{\mu x+(\mu^2+k^2)t},$$
$$G^{-}(x,t)=C^{\prime}e^{-\mu x-(\mu^2+k^2)t},$$
where $C$, $C^{\prime}$ are arbitrary constants.
Taking the integration constants as
$$C_1=e^{2\phi},\quad C_2=1,$$
$$C=e^{\phi^+},\quad C^{\prime}=-e^{\phi^-},$$
where $\phi$, $\phi^+$, $\phi^-$ satisfy
$$4k^2e^{2\phi}=e^{\phi^++\phi^-},$$
we obtain the solution of the RD system as
\begin{equation}
    \begin{aligned}
        \notag
        e^+&=\frac{e^{\mu x+(k^2+\mu^2)t+\phi^+}}
        {e^{k(x+2\mu t)+2\phi}+e^{-k(x+2\mu t)}},\\
        e^-&=-\frac{e^{-\mu x-(k^2+\mu^2)t+\phi^-}}
        {e^{k(x+2\mu t)+2\phi}+e^{-k(x+2\mu t)}}.\\
    \end{aligned}
\end{equation}
Note that this is actually the one-soliton solution of the RD system
\begin{equation}\label{1guzi}
    \begin{aligned}
        e^+&=\frac{e^{\eta^+}}{1+e^{\eta^++\eta^-+\phi_0}},\\
        e^-&=\frac{-e^{\eta^-}}{1+e^{\eta^++\eta^-+\phi_0}},\\
    \end{aligned}
\end{equation}
where
\begin{equation}
    \begin{aligned}
        \notag
        &e^{\phi_0}=\frac{1}{4k^2},\\
        &\eta^+=(\mu+k) x+(\mu+k)^2t+\phi^+,\\
        &\eta^-=-(\mu-k) x-(\mu-k)^2t+\phi^-.\\
    \end{aligned}
\end{equation}
Thus, by the transformation $(\ref{psigpm})$, the RNLS equation $(\ref{RNLS})$ has one-soliton solution
$$\Psi=re^{i\theta},$$
where
\begin{equation}
    \begin{aligned}
        \notag
        r=\frac{e^{\frac{\eta^++\eta^-}{2}}}{1+e^{\eta^++\eta^-+\phi_0}},\quad
        \theta=\frac{\eta^--\eta^+}{4}.
    \end{aligned}
\end{equation}

$\bullet$ two-soliton solution

Equation $(\ref{BRD})$ has a solution where
\begin{equation}
    \begin{aligned}
        \notag
        &F^{\prime}=e^{u_1+2\phi_1}+e^{-u_1},\\
        &G^{+\prime}=e^{v_1+\phi^+_1},\\
        &G^{-\prime}=-e^{-v_1+\phi^-_1},\\
    \end{aligned}
\end{equation}
with
\begin{equation}
    \begin{aligned}
        \notag
        &u_1=k_1x+2\mu_1 k_1t,\\
        &v_1=\mu_1 x+(\mu_1^2+k_1^2)t.
    \end{aligned}
\end{equation}

By substituting $(F^{\prime},G^{+\prime},G^{-\prime})$ as known solutions into $(\ref{Bäcklund})$, we obtain
\begin{equation}\label{2jieguzi}
    \begin{aligned}
        &G^+_{xx}+G^+_t-2k_1U_1G^+_x-2\mu_1k_1
        U_1G^+-(2\mu^2+k^2-k_1^2)G^+\\
        =&V_1^+[F_{xx}-F_t-2\mu_1 F_x-((2\mu^2+k^2)-(2\mu_1^2+k_1^2))F],\\
        &G^-_{xx}-G^-_t-2k_1U_1G^-_x+2\mu_1k_1
        U_1G^--(2\mu^2+k^2-k_1^2)G^-\\
        =&-V_1^-[F_{xx}+F_t+2\mu_1 F_x-((2\mu^2+k^2)-(2\mu_1^2+k_1^2))F],\\
        &F_{xx}-2k_1U_1 F_x-(k^2-k_1^2)F
        =V_1^-G^+-V_1^+G^-,\\
        &2\mu F_x-F_t-2k_1(\mu-\mu_1)U_1F
        =V_1^-G^++V_1^+G^-,\\
    \end{aligned}
\end{equation}
and we have
\begin{equation}
    \begin{aligned}
        \notag
        &G^+_x+V_1^+F_x=
        (\mu+k_1U_1)G^+
        -(\mu-\mu_1)V_1^+F,\\
        &G^-_x-V_1^-F_x=
        (-\mu+k_1U_1)G^-
        -(\mu-\mu_1)V_1^-F,\\
    \end{aligned}
\end{equation}
with
\begin{equation}
    \begin{aligned}
        \notag
        &U_1=\frac{e^{u_1+2\phi_1}-e^{-u_1}}{e^{u_1+2\phi_1}+e^{-u_1}},\\
        &V_1^+=\frac{e^{v_1+\phi^+_1}}{e^{u_1+2\phi_1}+e^{-u_1}},\\
        &V_1^-=\frac{e^{-v_1+\phi^-_1}}{e^{u_1+2\phi_1}+e^{-u_1}}.\\
    \end{aligned}
\end{equation}
Take the parameters $\mu=\mu_2,\ k=k_2$ and obtain the solution to equation $(\ref{BRD})$ as follows:
\begin{equation}\label{2guzi}
    \begin{aligned}
        F=&1+e^{\eta_1^++\eta_1^-+\phi_{01}}+e^{\eta_2^++\eta_2^-+\phi_{02}}\\
        &+\frac{e^{\eta_1^++\eta_2^-}}{(k_1+k_2+\mu_1-\mu_2)^2}+\frac{e^{\eta_1^-+\eta_2^+}}{(k_1+k_2-\mu_1+\mu_2)^2}\\
        &+(\frac{(k_1-k_2)^2-(\mu_1-\mu_2)^2}{(k_1+k_2)^2-(\mu_1-\mu_2)^2})^2
        e^{\eta_1^++\eta_1^-+\eta_2^++\eta_2^-+\phi_{01}+\phi_{02}},\\
    \end{aligned}
\end{equation}
\begin{equation}
    \begin{aligned}
        \notag
        G^+=&e^{\eta_1^+}+e^{\eta_2^+}+(\frac{k_1-k_2+\mu_1-\mu_2}{k_1+k_2-\mu_1+\mu_2})^2
        e^{\eta_1^++\eta_1^-+\eta_2^++\phi_{01}}\\
        &+(\frac{k_1-k_2+\mu_1-\mu_2}{k_1+k_2+\mu_1-\mu_2})^2e^{\eta_2^++\eta_2^-+\eta_1^++\phi_{02}},\\
        G^-=&-[e^{\eta_1^-}+e^{\eta_2^-}+(\frac{k_1-k_2-\mu_1+\mu_2}{k_1+k_2+\mu_1-\mu_2})^2
        e^{\eta_1^++\eta_1^-+\eta_2^-+\phi_{01}}\\
        &+(\frac{k_1-k_2-\mu_1+\mu_2}{k_1+k_2-\mu_1+\mu_2})^2e^{\eta_2^++\eta_2^-+\eta_1^-+\phi_{02}}],\\
    \end{aligned}
\end{equation}
where
\begin{equation}
    \begin{aligned}
        \notag
        &e^{\phi_{0i}}=\frac{1}{4k_i^2},\\
        &\eta_i^+=(\mu_i+k_i) x+(\mu_i+k_i)^2t+\phi^+_i,\\
        &\eta_i^-=-(\mu_i-k_i) x-(\mu_i-k_i)^2t+\phi^-_i.\\
    \end{aligned}
\end{equation}

It should be noted that this is a two-soliton solution for the bilinear RD system $(\ref{BRD})$. 
Similarly, the transformation $(\ref{psigpm})$ gives a two-soliton solution for the RNLS equation $(\ref{RNLS})$.

Generally, let
\begin{equation}
    \begin{aligned}
        \notag
        &F_0=1,\quad G_0^+=G_0^-=0,\\
        &F_{i_1i_2\cdots i_N}(x,t)=F(x,t;\ \mu_{i_1},k_{i_1};\cdots ;\ \mu_{i_N},k_{i_N}),\\
        &G^+_{i_1i_2\cdots i_N}(x,t)=G^+(x,t;\ \mu_{i_1},k_{i_1};\cdots ;\ \mu_{i_N},k_{i_N}),\\
        &G^-_{i_1i_2\cdots i_N}(x,t)=G^-(x,t;\ \mu_{i_1},k_{i_1};\cdots ;\ \mu_{i_N},k_{i_N}),\\
    \end{aligned}
\end{equation}
where $(F_{i_1i_2\cdots i_N},\ G^+_{i_1i_2\cdots i_N},\ G^-_{i_1i_2\cdots i_N})$ represents the solution of equation $(\ref{BRD})$ obtained 
by substituting $(F_{i_1i_2\cdots i_{N-1}},\ G^+_{i_1i_2\cdots i{N-1}},\ G^-_{i_1i_2\cdots i{N-1}})$ as the initial solution of 
equation $(\ref{BRD})$ in the Bäcklund transformation $(\ref{Bäcklund})$, and taking the parameters as
$$\mu=\mu_{i_N},\quad k=k_{i_N}.$$

It is shown in this subsection that $(F_1,\ G^+_1,\ G^-_1)$ and $(F_{12},\ G^+_{12},\ G^-_{12})$ are the one-soliton solution and the 
two-soliton solution of equation $(\ref{BRD})$, respectively.

Solving the double soliton solutions of the bilinear RD system $(\ref{BRD})$ using equation $(\ref{2jieguzi})$ involves dealing with 
a system of second-order partial differential equations, which is still quite complicated. The advantage of using the Bäcklund 
transformation to solve equations is that it allows for simplification of the solution process using nonlinear superposition formulas. 
For equation $(\ref{Bäcklund})$, notice that if substitut the initial solution
\begin{equation}
    \begin{aligned}
        \notag
        &F^{\prime}=e^{u_2+2\phi_2}+e^{-u_2},\\
        &G^{+\prime}=e^{v_2+\phi^+_2},\\
        &G^{-\prime}=-e^{-v_2+\phi^-_2},\\
    \end{aligned}
\end{equation}
with
\begin{equation}
    \begin{aligned}
        \notag
        &u_2=k_2x+2\mu_2 k_2t,\\
        &v_2=\mu_2 x+(\mu_2^2+k_2^2)t,
    \end{aligned}
\end{equation}
and take
$$\mu=\mu_1,\quad k=k_1,$$
Similarly, the bilinear RD system $(\ref{BRD})$ also has bilinear soliton solutions $(F_{21},\ G^+_{21},\ G^-_{21})$. Observing the 
bilinear soliton solutions $(\ref{2guzi})$ obtained in the previous section, under the permutation
$$k_1\mapsto k_2,\quad k_2\mapsto k_1,\quad \mu_1\mapsto \mu_2,\quad \mu_2\mapsto \mu_1$$
the form remains unchanged, indicating that the obtained solution is precisely $(F_{12},\ G^+_{12},\ G^-_{12})$, i.e., 
the bilinear transformation of the equation satisfies the nonlinear superposition formula
$$(F_{12},\ G^+_{12},\ G^-_{12})=(F_{21},\ G^+_{21},\ G^-_{21}).$$

In general, for any permutation $(i_1,i_2,\cdots,i_N)$ of $(1,2,\cdots,N)$, there is a nonlinear superposition formula
\begin{equation}\label{feixianxingdiejia}
    \begin{aligned}
        (F_{12\cdots N},\ G^+_{12\cdots N},\ G^-_{12\cdots N})=(F_{i_1i_2\cdots i_N},\ G^+_{i_1i_2\cdots i_N},\ G^-_{i_1i_2\cdots i_N}).
    \end{aligned}
\end{equation}

To apply the nonlinear superposition formula to the bilinear RD system $(\ref{BRD})$, the following properties of Hirota's 
bilinear operators need to be given.
\begin{lemma}\label{shuangxianxingprop2}
    $\textnormal{Hirota}$ operator satisfies
    \begin{equation}
        \begin{aligned}
            \notag
            (D_x^2a\cdot b)cd-ab(D_x^2c\cdot d)=D_x[(D_x a\cdot d)\cdot (bc)+(ad)\cdot (D_xc\cdot b)],
        \end{aligned}
    \end{equation}
    where $a$, $b$, $c$, $d$ are functions about $x$. 
\end{lemma}

For bilinear RD system $(\ref{BRD})$, by nonlinear superposition formula $(\ref{feixianxingdiejia})$, we have
\begin{theorem}\label{NS}
    Bilinear $\textnormal{RD}$ system $(\ref{BRD})$ has solution $F_{12\cdots N-1}$, $F_{12\cdots N}$, $F_{12\cdots N-1,N+1}$, $F_{12\cdots N+1}$
    satisfying nonlinear superposition formula
    \begin{equation}\label{feixianxingdiejiadairu}
        \begin{aligned}
            &2D_x[(D_xF_{12\cdots N}\cdot F_{12\cdots N-1,N+1})\cdot (F_{12\cdots N-1}F_{12\cdots N+1})]\\
            =&(F_{12\cdots N-1}F_{12\cdots N}G^-_{12\cdots N-1,N+1}+F_{12\cdots N-1}F_{12\cdots N-1,N+1}G^-_{12\cdots N})G^+_{12\cdots N+1}\\
            &+(F_{12\cdots N-1}F_{12\cdots N}G^+_{12\cdots N-1,N+1}+F_{12\cdots N-1}F_{12\cdots N-1,N+1}G^+_{12\cdots N})G^-_{12\cdots N+1}\\
            &-(G^+_{12\cdots N-1}G^-_{12\cdots N}F_{12\cdots N-1,N+1}+G^-_{12\cdots N-1}G^+_{12\cdots N}F_{12\cdots N-1,N+1}\\
            &+G^+_{12\cdots N-1}G^-_{12\cdots N-1,N+1}F_{12\cdots N}+G^-_{12\cdots N-1}G^+_{12\cdots N-1,N+1}F_{12\cdots N})F_{12\cdots N+1}.\\
        \end{aligned}
    \end{equation}
\end{theorem}

\begin{proof}
    By Bäcklund transformation$(\ref{Bäcklund})$ we have
    \begin{equation}\label{feixianxingdiejiajuti}
        \begin{aligned}
            &D_x^2F_{12\cdots N-1}\cdot F_{12\cdots N}=k_N^2F_{12\cdots N-1}F_{12\cdots N}-(G^+_{12\cdots N-1}G^-_{12\cdots N}+G^-_{12\cdots N-1}G^+_{12\cdots N}),\\
            &D_x^2F_{12\cdots N-1}\cdot F_{12\cdots N-1,N+1}=k_{N+1}^2F_{12\cdots N-1}F_{12\cdots N-1,N+1}\\
            &-(G^+_{12\cdots N-1}G^-_{12\cdots N-1,N+1}+G^-_{12\cdots N-1}G^+_{12\cdots N-1,N+1}),\\
            &D_x^2F_{12\cdots N}\cdot F_{12\cdots N+1}=k_{N+1}^2F_{12\cdots N}F_{12\cdots N+1}-(G^+_{12\cdots N}G^-_{12\cdots N+1}+G^-_{12\cdots N}G^+_{12\cdots N+1}),\\
            &D_x^2F_{12\cdots N-1,N+1}\cdot F_{12\cdots N+1}=k_N^2F_{12\cdots N-1,N+1}F_{12\cdots N+1}\\
            &-(G^+_{12\cdots N-1,N+1}G^-_{12\cdots N+1}+G^-_{12\cdots N-1,N+1}G^+_{12\cdots N+1}).\\
        \end{aligned}
    \end{equation}

    substitute $(\ref{feixianxingdiejiajuti})$ to
    $$k_N^2F_{12\cdots N-1}F_{12\cdots N}F_{12\cdots N-1,N+1}F_{12\cdots N+1}$$
    and 
    $$k_{N+1}^2F_{12\cdots N-1}F_{12\cdots N-1,N+1}F_{12\cdots N}F_{12\cdots N+1}$$
    respectively, we have
    \begin{equation}
        \begin{aligned}
            \notag
            &[D_x^2F_{12\cdots N-1}\cdot F_{12\cdots N}+H_1]F_{12\cdots N-1,N+1}F_{12\cdots N+1}\\
            =&F_{12\cdots N-1}F_{12\cdots N}[D_x^2F_{12\cdots N-1,N+1}\cdot F_{12\cdots N+1}+H_2],\\
            &[D_x^2F_{12\cdots N-1}\cdot F_{12\cdots N-1,N+1}+H_3]F_{12\cdots N}F_{12\cdots N+1}\\
            =&F_{12\cdots N-1}F_{12\cdots N-1,N+1}[D_x^2F_{12\cdots N}\cdot F_{12\cdots N+1}+H_4],\\
        \end{aligned}
    \end{equation}
    with
    \begin{equation}
        \begin{aligned}
            \notag
            &H_1=G^+_{12\cdots N-1}G^-_{12\cdots N}+G^-_{12\cdots N-1}G^+_{12\cdots N},\\
            &H_2=G^+_{12\cdots N-1,N+1}G^-_{12\cdots N+1}+G^-_{12\cdots N-1,N+1}G^+_{12\cdots N+1},\\
            &H_3=G^+_{12\cdots N-1}G^-_{12\cdots N-1,N+1}+G^-_{12\cdots N-1}G^+_{12\cdots N-1,N+1},\\
            &H_4=G^+_{12\cdots N}G^-_{12\cdots N+1}+G^-_{12\cdots N}G^+_{12\cdots N+1}.\\
        \end{aligned}
    \end{equation}

    By lemma $\ref{shuangxianxingprop2}$ we finish the proof.

\end{proof}
The equation $(\ref{feixianxingdiejiadairu})$ and the Bäcklund transformation $(\ref{Bäcklund})$ can be combined to obtain 
$(F_{12\cdots N+1},\ G^+_{12\cdots N+1},\ G^-_{12\cdots N+1})$. The difficulty of solving the combined equations is lower than 
that of the Bäcklund transformation itself. We apply Theorem $\ref{NS}$ to solve the bilinear RD system $(\ref{BRD})$ for the 
double solitary wave solution, and use this to demonstrate the specific process of simplifying the equation.

Taking the solutions $(F_1,\ G^+_1,\ G^-_1)$ and $(F_2,\ G^+_2,\ G^-_2)$ of equation $(\ref{BRD})$, theorem $\ref{NS}$ yields
\begin{equation}
    \begin{aligned}
        \notag
        &2D_x[(D_xF_1\cdot F_2)\cdot(F_0F_{12})]=F_0(F_1G^+_1+F_2G^+_2)G^-_{12}+F_0(F_1G^-_1+F_2G^-_2)G^+_{12},\\
    \end{aligned}
\end{equation}
Thus for $(F_{12},\ G^+_{12},\ G^-_{12})$ we have
\begin{equation}\label{2jiefeixianxingdiejia}
    \begin{aligned}
        F_{12x}-\frac{(F_{1x}F_2-F_1F_{2x})_x}{F_{1x}F_2-F_1F_{2x}}F_{12}
        +\frac{F_1G^+_1+F_2G^+_2}{F_{1x}F_2-F_1F_{2x}}G^-_{12}+\frac{F_1G^-_1+F_2G^-_2}{F_{1x}F_2-F_1F_{2x}}G^+_{12}=0.
    \end{aligned}
\end{equation}

We can verify that $(\ref{2guzi})$ is solution to eqaution.
Note that by applying Theorem $\ref{NS}$, the second-order partial differential equation $(\ref{2jieguzi})$ can be simplified to the 
above first-order partial differential equation system, in the form of
\begin{equation}
    \begin{aligned}
        \notag
        &\partial_x(F,G^+,G^-)=L_1(F,G^+,G^-),\\
        &\partial_t(F,G^+,G^-)=L_2(F,G^+,G^-),\\
    \end{aligned}
\end{equation}
where $L_1$ and $L_2$ are homogeneous linear functions. This is actually two first-order linear homogeneous ordinary differential 
equation systems, which greatly reduces the difficulty of solving the equation $(\ref{2jieguzi})$.

In general, to solve the $N$-soliton solution of equation $(\ref{BRD})$, we can substitute the $(N-1)$-soliton solution into the
Bäcklund transformation $(\ref{Bäcklund})$ for solution, or solve it through nonlinear superposition formula with two different 
parameter $(N-1)$-soliton solutions.

\section{Conclusion}

This paper presents a weak reduction condition and provides rational solutions for the nonlocal RNLS equation under this condition. 
The specific verification of the weak reduction condition is not conducted in this paper. However, by applying this condition, 
general high-order rational solutions for the RNLS equation are derived, as well as first-order rational solutions for the nonlocal 
RNLS equation. This demonstrates that the weak reduction condition proposed in this paper is meaningful for both the nonlocal case 
and the case of high-order rational solutions. It should be noted that the simplification using Schur polynomials, which is employed 
for high-order rational solutions of the RNLS equation in this paper, has its limitations and does not hold for nonlocal RNLS equations. 
Further research can be conducted to investigate the general high-order rational solutions of nonlocal RNLS equations, which relies 
on the specific verification of the weak reduction condition.
We have also studied the solitary wave solutions of the RNLS equation using the Bäcklund transformation. We obtained one-
soliton solutions and two-soliton solutions, and simplified the process of solving the double soliton solutions using nonlinear 
superposition formulas.

\section*{Acknowledgments}
The work was supported in part by the National Natural Science Foundation of China (No. 12171098, 11571079, 11701322), The Natural Science Foundation of Shanghai (No. 14ZR1403500), Shanghai Pujiang Program (No. 14PJD007).

\end{document}